\documentclass[prb,twocolumn,superscriptaddress,floatfix,nobibnotes]{revtex4}
\usepackage{color,graphicx}
\usepackage{subfigure}
\usepackage{amsmath,amsfonts,amssymb,amsthm,verbatim}
\usepackage{tabularx}
\usepackage{bm}
\allowdisplaybreaks[4]
\def\be{\begin{equation}}
\def\ee{\end{equation}}
\def\ba{\begin{eqnarray}}
\def\ea{\end{eqnarray}}

\newcommand{\br}{\mathbf{r}}

\newcommand{\bk}{\mathbf{k}}

\begin{document}
\title{Hedgehog spin texture and competing orders associated with strains on the surface of a topological crystalline insulator}
\author{Cheng-Yi Huang}
\affiliation{Department of Physics, National Sun Yat-sen University, Kaohsiung 80424, Taiwan}
\author{Hsin Lin}
\email{nilnish@gmail.com}
\affiliation{Centre for Advanced 2D Materials and Graphene Research Centre, National University of Singapore, Singapore 117546}
\affiliation{Department of Physics, National University of Singapore, Singapore 117542}
\author{Yung Jui Wang}
\affiliation{Department of Physics, Northeastern University, Boston, Massachusetts 2115,USA}
\author{Arun Bansil}
\affiliation{Department of Physics, Northeastern University, Boston, Massachusetts 2115,USA}
\author{Wei-Feng Tsai}
\email{wftsai@mail.nsysu.edu.tw}
\affiliation{Department of Physics, National Sun Yat-sen University, Kaohsiung 80424, Taiwan}
\date{\today}
\begin{abstract}
%The spotlight of our work: (1) Spin texture due to strain; (2) tunable Chern number via interplay between strain and Zeeman field; (3) symmetry breaking orders via short-range repulsion in the absence/presence of strain.

We have investigated spin reorientation phenomena and interaction driven effects under the presence of applied strains on the (001) surface of Pb$_{1-x}$Sn$_x$(Te, Se) topological crystalline insulators, which host multiple Dirac cones. Our analysis is based on a four-band $k\cdot p$ model, which captures the spin and orbital textures of the surface states at low energies around the $\bar{X}$ and $\bar{Y}$ points, including the Lifshitz transition. Even without breaking the time-reversal symmetry, we find that certain strains which break the mirror symmetry can induce hedgehog-like spin texture associated with gap formation at the Dirac points. The Chern number of the gapped surface ground state is shown to be tunable through the interplay of strains and a perpendicular Zeeman field.  We also consider effects of strain in the presence of interactions in driving competing orders, and obtain the associated phase diagram at the mean-field level. Potential applications of our results for low power consuming electronics are discussed. 
\end{abstract}
\maketitle

\section{Introduction}
%(1) What is TCI: a new state of matter. 
A topological insulating (TI) phase is a new state of quantum matter featuring massless Dirac-like boundary states whose robustness is guaranteed by the time-reversal symmetry (TRS)\cite{Hasan10,Qi11,Moore10,Ando13}. In this rapidly growing field, search for new topological phases has recently turned to extending the consideration of symmetry-protected states to include non-spatial symmetries\cite{Schnyder08,Kitaev09} such as, the particle-hole and chiral symmetries as well as the spatial symmetries\cite{Hughes10,Turner12,Fang12,Slager13,Sato14}. In particular, the nontrivial band topology can be shown to be protected by certain crystal symmetries, leading to the new class of TIs called topological crystalline insulators (TCIs)\cite{LFTCI,mChen,Chiu13}. Like the TIs, a three dimensional (3D) TCI is also predicted to host metallic surface states on surfaces which preserve appropriate crystal symmetries\cite{Hsieh12}.

%(2) Experimental status. The presence of strain.

To date, the semiconducting Pb$_{1-x}$Sn$_x$(Te, Se) alloys are the only experimentally realized TCI materials, whose topological nature has been verified through direct observation of Dirac-like surface states via angle-resolved photoemission spectroscopy (ARPES) experiments\cite{Xu12a,Dziawa12,Tanaka12}. These gapless surface states are protected by the mirror symmetry with respect to the (110) or ($\bar{1}$10) lattice plane. The characteristic features of the associated Dirac states such as linear dispersion, Lifshitz transition, spin/orbital texture, among others, have been examined via scanning tunneling microscopy/spectroscopy (STM/STS) and transport measurements\cite{Xu12a,Yazdani13,Taskin14,OkadaTCIstrain,ZeljkovicTCIphase,ZeljkovicTCIorbital}. Moreover, it has been found that, below a critical temperature, which depends on chemical composition, the cubic lattice structure can be distorted by strains, resulting in an orthorhombic or a rhombohedral structure\cite{Iizumi75}. Remarkably, in a recent STM experiment by Okada {\it et al.}\cite{OkadaTCIstrain}, two unexpected Landau levels have been observed for the (001) surface of Pb$_{1-x}$Sn$_x$Se under a perpendicular magnetic field. The presence of these two extra levels is believed to be associated with the gap opening of the two surface Dirac cones induced by a ferroelectric-like lattice distortion, which breaks the corresponding mirror symmetry. It is clear that it is important to understand effects of strains on the electronic structure of TCIs in order to gain a deeper handle on the nature and origin of their topological states, and how these states could be manipulated for practical applications of the TCIs.

%(3) Our goal: control spin d.o.f. of the surface states from two ways, i) 
%strain and ii) electron-electron interactions.

Despite some strain-related studies in the literature\cite{Fang14,Serbyn14}, we are not aware of a systematic investigation of how spin-textures and other properties of Dirac states in the TCIs evolve under various symmetry breaking strains. So motivated, here we examine strain effects on the (001) surface of the Pb$_{1-x}$Sn$_x$(Te, Se) TCIs. Our analysis is based on an effective four-band model, which is shown to capture all essential features of the topological surface states. The form of strain related perturbations on the Hamiltonian is clarified through general symmetry considerations, allowing us to delineate how characteristic features of the topological states in TCIs evolve under strains.

Significantly, our analysis shows that certain mirror-symmetry-breaking strains induce hedgehog spin texture with out-of-the-plane spin-polarization at the Dirac point. Such a spin texture of Dirac states has been reported previously in a 3D TI, but only in the presence of an exchange field\cite{Xu12b}. Our findings thus identify a possible new pathway for realizing TCI-based spintronics devices without requiring a TRS breaking field\cite{Liu14}. Moreover, we show that the interplay of strains and applied Zeeman field can be used to tune the Chern number of the surface ground state, which is a topological invariant characterizing a quantum anomalous Hall (QAH) or a quantum Hall insulator.

Electron-electron interactions effects are ubiquitous in condensed matter systems, especially when charge screening is relatively poor as is the case generally for Dirac-like states in graphene\cite{CastroNeto09,Wehling11} or TIs and TCIs.\cite{Coulomb} In this connection, we discuss possible symmetry breaking orders generated under short-range repulsion $U$ as a first step toward understanding correlation driven effects on Dirac states in the presence of applied strains. We numerically obtain the zero-temperature phase diagram as a function of $U$ and the particle density $n$ and delineate its evolution with increasing strength of strain. Our results not only give insight into the nature of competing orders, but also build the foundation for what may be called ``straintronics'' applications driven by electron correlations.

%giving hints for possible applications.

%(4) Paper organization.

The paper is organized as follows. In Sec. II, we briefly discuss the effective four-band model for the (001) surface states in the low-energy regime, and its extensions for addressing strain effects in TCIs. Sec. III turns to consider strain effects on properties of the Dirac states, including the interplay between applied strains and perpendicular Zeeman fields. These results then allow us to investigate in Section IV the possible interaction-driven competing orders in Dirac states in the absence as well as presence of strains. Finally, Section V comments on potential applications and implications of our study, and concludes with a summary of our results.

\section{Effective four-band $k\cdot p$ model for the surface states}
\begin{figure}[t]
\begin{center}
\includegraphics[width=8cm]{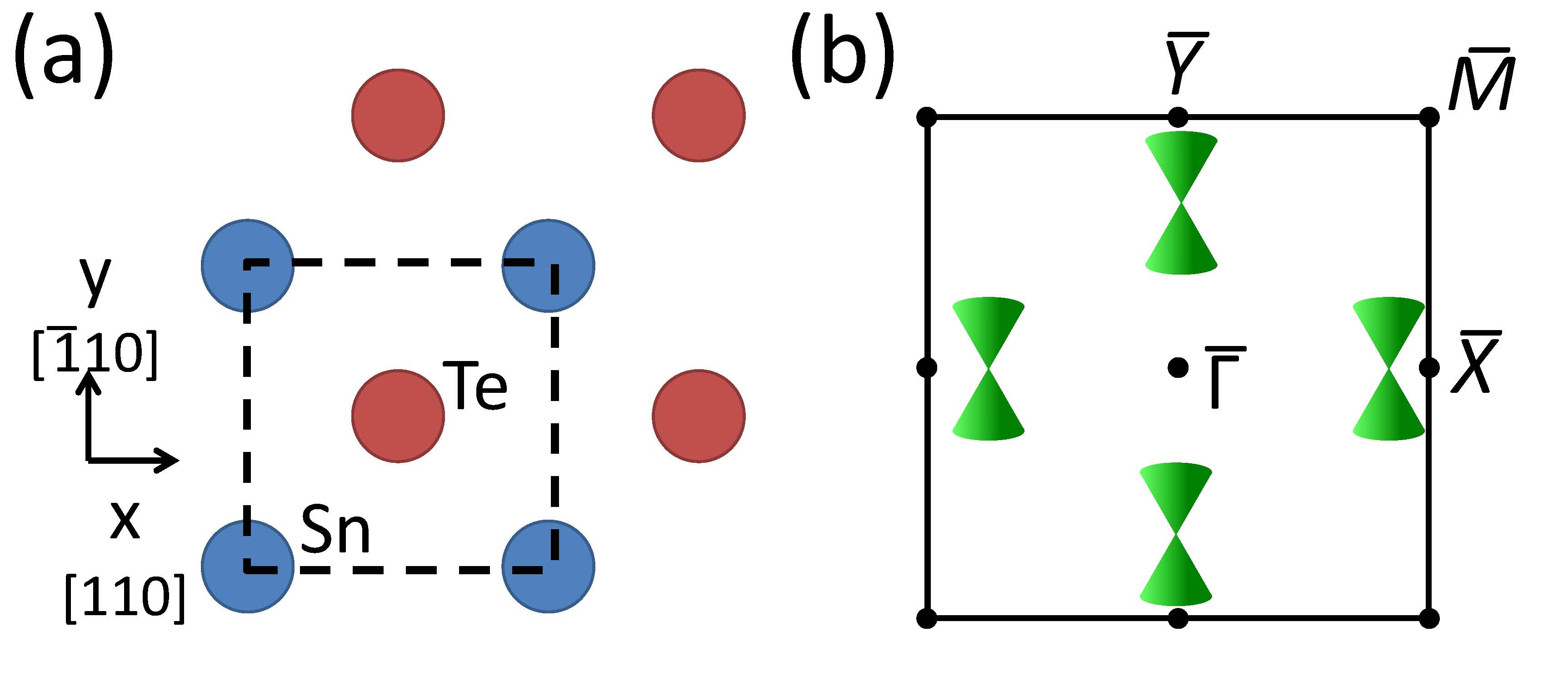}
\caption{(Color online) (a) Top view of the crystal structure of SnTe. Two types of atoms are drawn in different colors. $x$ ($y$) denotes the direction along [110] ([$\bar{1}$10]). (b) Surface Brillouin zone of SnTe. Without strain, there are four gapless Dirac cones, protected by the mirror symmetry either with respect to $\bar{\Gamma}\bar{X}$ or $\bar{\Gamma}\bar{Y}$.}
\label{fig:lattice}
\end{center}
\end{figure}

%Introduce k.p model and symmetries
We start by reviewing the four-band $k\cdot p$ model for the (001) surface states in a TCI developed previously in Ref.\onlinecite{PhysRevB.87.235317}, and discuss its generalization to account for effects of strains. On the (001) surface, the low-energy surface states can be viewed as two sets of interacting coaxial Dirac cones, originating from the interface between the inverted bands of the TCI and the vacuum: one set is centered at $\bar{X}$ and the other at $\bar{Y}$ in the surface Brillouin zone (SBZ), see Fig.~\ref{fig:lattice} (b). Defining [110], [$\bar{1}$10], and [001] as $x$, $y$, and $z$ directions [see Fig.~\ref{fig:lattice} (a)], respectively, the effective model $H_{\bar{X}}$ around $\bar{X}$ must obey the following three essential symmetries which leave $\bar{X}$ invariant: the mirror reflection about the $xz$-plane, the mirror reflection about the $yz$-plane, and time-reversal symmetry with the corresponding symmetry operations, i.e. 
\ba
M_{xz}H_{\bar{X}}(k_x,k_y)M_{xz}^{-1}=H_{\bar{X}}(k_x,-k_y),\\
M_{yz}H_{\bar{X}}(k_x,k_y)M_{yz}^{-1}=H_{\bar{X}}(-k_x,k_y),\\
TH_{\bar{X}}(k_x,k_y)T^{-1}=H_{\bar{X}}(-k_x,-k_y),
\ea
where $\bk=(k_x,k_y)$ is measured from $\bar{X}$.

Informed by the orbital characters of surface states as revealed by first-principles calculations,\cite{PhysRevB.87.235317} one may choose, for instance in the case of SnTe, $\{|p_z,\uparrow;\text{Sn}\rangle$, $
|p_z,\downarrow;\text{Sn}\rangle$, $|p_x,\uparrow;\text{Te}\rangle$, $
|p_x,\downarrow;\text{Te}\rangle\}$ as the basis states of $H_{\bar{X}}$ because one of the interacting Dirac cones is $p_z$-orbital (Sn) rich, while the other is $p_x$-orbital (Te) rich. As a result, the symmetry operations can be represented by the matrices, $M_{xz}=-i\Sigma_{02}$, $M_{yz}=-i\Sigma_{31}$, and $T=-i\Sigma_{02}K$, where $\Sigma_{\alpha\beta}\equiv\sigma_{\alpha}\otimes s_{\beta}$ with the Pauli matrices $\vec{\sigma}$ and $\vec{s}$ acting on orbital and spin spaces, respectively, and $K$ denotes complex conjugation. These considerations restrict $H_{\bar{X}}$ to the form
\ba
H_{\bar{X}}&=& m\Sigma_{30}+m^\prime\Sigma_{22}+k_x(v_{1x}\Sigma_{02}+v_{2x}
\Sigma_{20}+v_{3x}\Sigma_{32})\nonumber \\
&+& k_y(-v_{1y}\Sigma_{01}+v_{2y}\Sigma_{13}-v_{3y}\Sigma_{31}),\label{H}
\label{eq:surfaceH}
\ea
up to first order in $|\bk|$.

%NOTE: lettering in many figures should be larger, and could be made similar to that in Figs. 6, 8 and 9, for example. ARUN

%NOTE: The unit cell should perhaps be drawn in the crystal structure Fig 1(a) and other similar figures later--arun

%Model structure: parent cone (parameters off) and Ed1, Ed2; Lifshitz points; %child cones; zero sz; fitting parameters.

\begin{figure*}[th]
\begin{center}
\includegraphics[width=18cm]{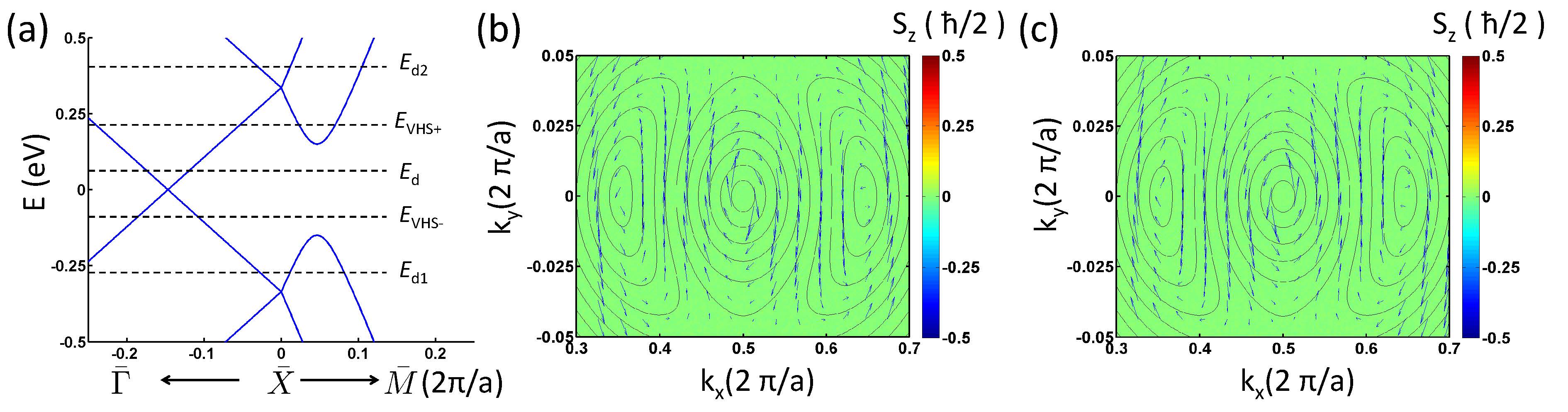}
\caption{(Color online) (a) Band dispersion of the (001) surface of SnTe along high symmetry lines obtained by setting model parameters in Eq.~(\ref{eq:surfaceH}) as: $a=6.327$ \AA, $m = -0.3$ eV,  $m^\prime= -0.15$ eV, $v_{2x} = v_{2y} = v_{3x} = v_{3y} = 0$ eV\AA, $v_{1x} = -2.3$ eV\AA, and $v_{1y} = -6.5$ eV\AA. $E_d$ and $E_{d1(2)}$ denote energies of the low-energy Dirac points and the two associated parent Dirac points, respectively. $E_{VHS+(-)}$ are energies of the van Hove singularities. (b) and (c) separately show spin textures of valence and conduction bands. Arrows and their lengths indicate directions and magnitudes of the in-plane spin polarization; the color map indicates the distribution of the out-of-the-plane spin polarization. Solid lines mark contours of constant energy.}
\label{fig:b1}
\end{center}
\end{figure*}

For $m^\prime$=0 and $v_{2x}$=$v_{3x}$=$v_{2y}$=$v_{3y}$=0, $H_{\bar{X}}$ becomes block diagonal, reflecting the structure of the two underlying parent Dirac cones with two associated Dirac points, $E_{d1}=-m$ ($p_z$-rich) and $E_{d2}=m$ ($p_x$-rich) at $\bar{X}$ [see Fig.~\ref{fig:b1} (a)]. A non-vanishing value of any of the parameters $m'$, $v_{2x}$, $v_{3x}$, $v_{2y}$ and $v_{3y}$ induces interaction between the two cones, leading to two immediate consequences: (1) The original Dirac points are shifted to $E_{d1(2)}=\mp\sqrt{m^2+m^{\prime 2}}$; and (2) All degeneracies along the intersection of the two cones are lifted except for the two points of a time-reversal symmetric pair on the $\bar{\Gamma}\bar{X}$ line, indicating the emergence of two child Dirac cones protected by the mirror reflection about the $xz$-plane [see Fig.~\ref{fig:lattice} (b)]; hereafter, referred to as the low-energy Dirac points associated with $E_d$=0.

Our effective four-band $k\cdot p$ model correctly capture two key features of surface states of TCIs. Firstly, for $E>0$ topology of the constant energy contours changes from two separate Fermi circles at low energies to two concentric ellipses at high energies. This Lifshitz transition\cite{Hsieh12} indicates the presence of van Hove singularities (VHSs) in the underlying electronic spectrum, which are located along the $\bar{X}\bar{M}$ line at 
\be
E_{VHS+} = |m^{\prime}|,\quad \bk = (0,\pm\frac{m}{v_{1y}}),
\ee
if $v_{2x}$, $v_{3x}$, $v_{2y}$, and $v_{3y}$ are neglected. A similar situation arises for $E<0$ with $E_{VHS-}=-E_{VHS+}$. In all cases, we have a logarithmically diverging density of states proportional to $\ln \frac{\Lambda}{|\omega|}$ with $\omega$ and $\Lambda$ representing an energy scale away from the VHS and a cutoff energy for surface states, respectively. Secondly, there is no out-of-the-plane spin component [Fig.~\ref{fig:b1} (b)-(c)] by using symmetry arguments in that the net out-of-the-plane spin polarization, $\langle\Sigma_{03}\rangle=\langle M_{xz}M_{yz}T\Sigma_{03}T^{-1}M_{yz}^{-1}M_{xz}^{-1}\rangle=-\langle \Sigma_{03}\rangle$, yielding  $\langle\Sigma_{03}\rangle =0$.

We turn next to generalize our four-band effective model to include low-energy states throughout the SBZ in the presence of inter-cone interactions. This can be done by noticing that the surface states in the vicinity of $\bar{X}$ and $\bar{Y}$ are related by $C_4$ symmetry, so that the effective model around $\bar{Y}$ can be obtained explicitly by a $C_4$ rotation,
\be 
H_{\bar{Y}}(k_x,k_y)=\hat{C}_4 H_{\bar{X}}(k_y,-k_x)\hat{C}_4^{-1},
\ee
where $\hat{C}_4=\sigma_0\otimes e^{-i\frac{\pi}{4}s_3}$ and the $p_x$ orbital for Te atoms in the basis states is now replaced by $p_y$ orbital in $H_{\bar{Y}}$. The total surface Hamiltonian then is 
\be 
H_{(001)}=H_{\bar{X}}(k_x,k_y)\oplus H_{\bar{Y}}(k_x,k_y).
\ee
Although $H_{(001)}$ describes an even number of Dirac cones at low energies like the case of a weak TI\cite{Fu07a,Fu07b,LFTCI2,Yan12}, note that the four Dirac points in a TCI do not locate at time-reversal invariant momenta. This key difference can lead to a rather different phase diagram in the presence of the electron-electron interactions\cite{Liu2012906} in a TCI, as discussed in Sec. IV below.

\section{STRAIN EFFECTS}
%Derivation of the strain-induced perturbation terms (Breaking symmetries with %corresponding to strain tensor)

\subsection{Gap opening and spin textures}
We now discuss effects on the surface states due to strains, which could be either intrinsic or extrinsic. Since the gapless surface states in the Pb$_{1-x}$Sn$_x$(Te,Se) TCIs are mainly protected by mirror symmetries, a perturbation, which breaks one of these symmetries can be expected to at least partly gap the surface spectrum, and modify the spin textures around the associated (massive) Dirac points. Although this is generally true, we will see below that this is not always the case due to other symmetry considerations. For this purpose, we will carry out a systematic analysis of strain induced effects along the lines of Sec. II above. 

\begin{figure*}[th]
\begin{center}
\includegraphics[width=18cm]{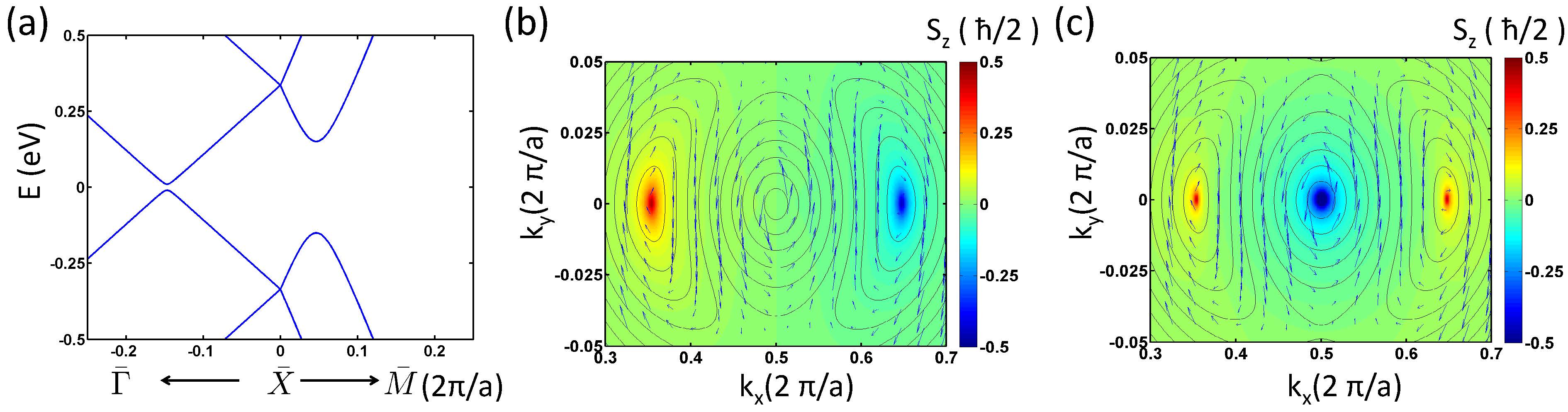}
\caption{(Color online) (a) Band dispersion of (001) surface of SnTe along high symmetry lines with the strain perturbation $\Delta\Sigma_{23}$; parameters are the same as those used in Fig.~\ref{fig:b1} and $\Delta=0.01$ eV. (b) and (c) are spin textures of the conduction band in the presence of the strain and a perpendicular Zeeman field [modeled by adding $H^\prime=h_z\Sigma_{03}$ in Eq.~(\ref{eq:surfaceH}) with $h_z=0.01$ eV], respectively. Arrows and their lengths indicate the direction and the magnitude of in-plane spin polarizations; color maps show the distribution of the out-of-the-plane spin polarization. Solid lines mark constant energy contours.}
\label{fig:f1}
\end{center}
\end{figure*}

A general strain can be described in terms of a symmetric strain tensor $\varepsilon_{ij}=\partial_j u_i$, where $\mathbf{u}$ denotes the displacement field and $i,j=x,y,z$ are chosen to coincide with the principal crystal axes as defined in Sec. II above. Any effective, strain-induced perturbation term corresponding to $\varepsilon_{ij}$ will transform in the same way as the strain tensor under time-reversal and mirror symmetries. By examining the transformation properties of $\Sigma_{\alpha\beta}$ up to first order in $|\bk|$, the possible resulting perturbations around $\bar{X}$ are listed in Table \ref{T1}.

\begin{table*}
\caption{Effects of perturbations induced by applied strains. The first row shows the form of each perturbation in terms of a $4\times 4$ matrix in the basis of the effective model around $\bar{X}$. $\Delta$ and $v$ are the coupling strengths. The next two rows indicate if a perturbation breaks certain mirror symmetry. The fourth and fifth rows represent the possible consequences. The last row shows the corresponding strain tensor with which each perturbation can couple.}
\begin{ruledtabular}
\begin{tabular}{ccccccccccccc}
Perturbation &$\Delta\Sigma_{23}$&$vk_x\Sigma_{03}$&$vk_x\Sigma_{11}$&$vk_x\Sigma_{33}$
&$\Delta\Sigma_{10}$&$vk_x\Sigma_{12}$&$\Delta\Sigma_{21}$& $vk_x\Sigma_{01}$&$vk_x\Sigma_{13}$&$vk_x\Sigma_{31}$ \\
Break $xz$-mirror symmetry
&$\checkmark$&$\checkmark$&$\checkmark$&$\checkmark$&$\times$&$\times$
&$\checkmark$&$\checkmark$&$\checkmark$&$\checkmark$ \\
Break $yz$-mirror symmetry
&$\times$&$\times$&$\times$&$\times$&$\checkmark$&$\checkmark$&$\checkmark$
&$\checkmark$&$\checkmark$&$\checkmark$ \\
Open gaps at Dirac points
&$\checkmark$&$\checkmark$&$\checkmark$&$\checkmark$&$\times$&$\times$&$\times$
&$\times$&$\times$&$\times$ \\
Induce out-of-plane spins
&$\checkmark$&$\checkmark$&$\checkmark$&$\checkmark$&$\times$&$\times$&$\times$
&$\times$&$\times$&$\times$ \\
Coupled strain tensor
&$\varepsilon_{yz}$&$\varepsilon_{yz}$&$\varepsilon_{yz}$&$\varepsilon_{yz}$
&$\varepsilon_{xz}$&$\varepsilon_{xz}$&$\varepsilon_{xy}$&$\varepsilon_{xy}$
&$\varepsilon_{xy}$&$\varepsilon_{xy}$
\label{T1}
\end{tabular}
\end{ruledtabular}
\end{table*}

As expected, the strain induced perturbations in the first four columns of Table I can open gaps at both low-energy Dirac points along $\bar{\Gamma}\bar{X}$ due to $M_{xz}$ broken symmetry. More significantly, the resulting spin texture becomes hedgehog-like at low energies, similar to the spin reorientation phenomenon found on the surfaces of manganese-doped Bi$_2$Se$_3$ thin films\cite{Xu12b}.

As a concrete example, let us add the perturbation $\Delta\Sigma_{23}$ in Eq.(\ref{eq:surfaceH}), where $\Delta$ denotes the electron-phonon coupling strength. Clearly, in Fig.~\ref{fig:f1}(a), a gap opens at the Dirac point with its magnitude proportional to $|\Delta|$. The spin-texture of the conduction band of the massive Dirac cone is also depicted in Fig.~\ref{fig:f1}(b). As the absolute value of the eigenenergy $|E|$ becomes smaller ({\it i.e.}, closer to the bottom of the upper cone), the induced out-of-the-plane spin component,
\be
\langle\Sigma_{03}\rangle
=\frac{2m'\Delta v_{1x}k_x}{E[E^2-(m^2+m'{}^2+v_{1x}^2k_x^2+v_{1y}^2k_y^2+\Delta^2)]},
\label{eq:sz}
\ee
becomes larger, where we have set $v_{2x}=v_{2y}=v_{3x}=v_{3y}=0$ for simplicity. Several points should be noted here as follows. Firstly, since the strain perturbation does not break TRS, the spin texture for the two massive cones around $\bar{X}$ must form time-reversed partners. This can be inferred from Eq.~(\ref{eq:sz}) by changing the sign of $k_x$. Secondly, the out-of-the-plane spin component of the lower cone is basically opposite to that of the upper cone at the same $|\bk|$. Finally, the whole spin texture resulting from the strain differs from the case where the hedgehog-like texture is induced by the perpendicular Zeeman field when both massive Dirac cones around $\bar{X}$ are considered, see Fig.~\ref{fig:f1}(c).

Other strain perturbations in Table \ref{T1}, which break either the $yz$-mirror symmetry or both the mirror symmetries, neither open a gap at the Dirac point nor induce an out-of-the-plane spin component. The robustness of this gapless Dirac point originates from an underlying symmetry: the former type of perturbation is due to the presence of $M_{xz}$ symmetry, while the latter type (shear deformation) is protected by a ``space-time'' symmetry, $C_{2T}=i\Sigma_{31}K$, with a rotation $C_2$ followed by a time-reversal operation. The space-time protection now allows persistence of gapless Dirac points sitting at generic $\bk$ points away from the mirror line $\bar{\Gamma}\bar{X}$. This special feature has been noted previously in Refs.\onlinecite{Fang14,Serbyn14}, and it leads to a new type of TCI\cite{Fang15}.

Vanishing out-of-the-plane spin polarization can be proven via symmetry arguments. For instance, consider the surface states around $\bar{X}$ under the strain perturbation $\Delta\Sigma_{10}$. The corresponding Hamiltonian now reads: $\tilde{H}_{\bar{X}} = H_{\bar{X}}+\Delta\Sigma_{10}$. Assuming $|\Psi\rangle$ is an eigenstate of $\tilde{H}_{\bar{X}}$, the $xz$-mirror symmetry guarantees that
\be
0=\langle\Psi|[\Sigma_{02},\tilde{H}_{\bar{X}}]|\Psi\rangle=2iv_{1y}k_y
\langle\Psi|\Sigma_{03}|\Psi\rangle.
\ee
Thus, the out-of-the-plane spin component $\langle\Psi|\Sigma_{03}|\Psi\rangle$ vanishes everywhere around $\bar{X}$. Other perturbations can be analyzed in a similar manner.

There other cases deserve some comment. The first two cases involve a uniform expansion and a uniaxial stretch ($C_4$ breaking), which correspond to $\varepsilon_{xx}+\varepsilon_{yy}$ and $\varepsilon_{xx}-\varepsilon_{yy}$, respectively\cite{Zeljkovic15}. Because both these cases respect TRS as well as the mirror symmetries, their net effect is only to renormalize the parameters in the original $H_{\bar{X}}$. Consequently, these perturbations do not open a gap at Dirac points, although positions of the Dirac points could shift in opposite directions along the $\bar{\Gamma}\bar{X}$ line. The third case is an experimentally observed ferroelectric-like distortion\cite{OkadaTCIstrain}, in which two kinds of atoms are displaced along a certain direction in an opposite manner. Denoting the displacement vector $\textbf{d}=(d_x,d_y)$, the non-vanishing component of this distortion, $d_x$ ($d_y$), preserves the mirror symmetry with respect to the principal axis $x$ ($y$), but breaks the rotation and the mirror symmetries along a perpendicular direction. To the zeroth order in $\bk$ around $\bar{X}$, the perturbation due to such distortion can be straightforwardly shown to have the form: 
\be
H_F^\prime = \Delta_{Fx}d_x\Sigma_{10}+\Delta_{Fy}d_y\Sigma_{23},
\ee
where $\Delta_{Fx}$ and $\Delta_{Fy}$ denote the coupling strengths. In fact, the former term is similar to the effect of strain $\varepsilon_{xz}$, preserving the gapless Dirac points, while the latter term is similar to the effect of strain $\varepsilon_{yz}$, resulting in opening a gap instead.

\begin{figure}[tbh]
\begin{center}
\includegraphics[width=7cm]{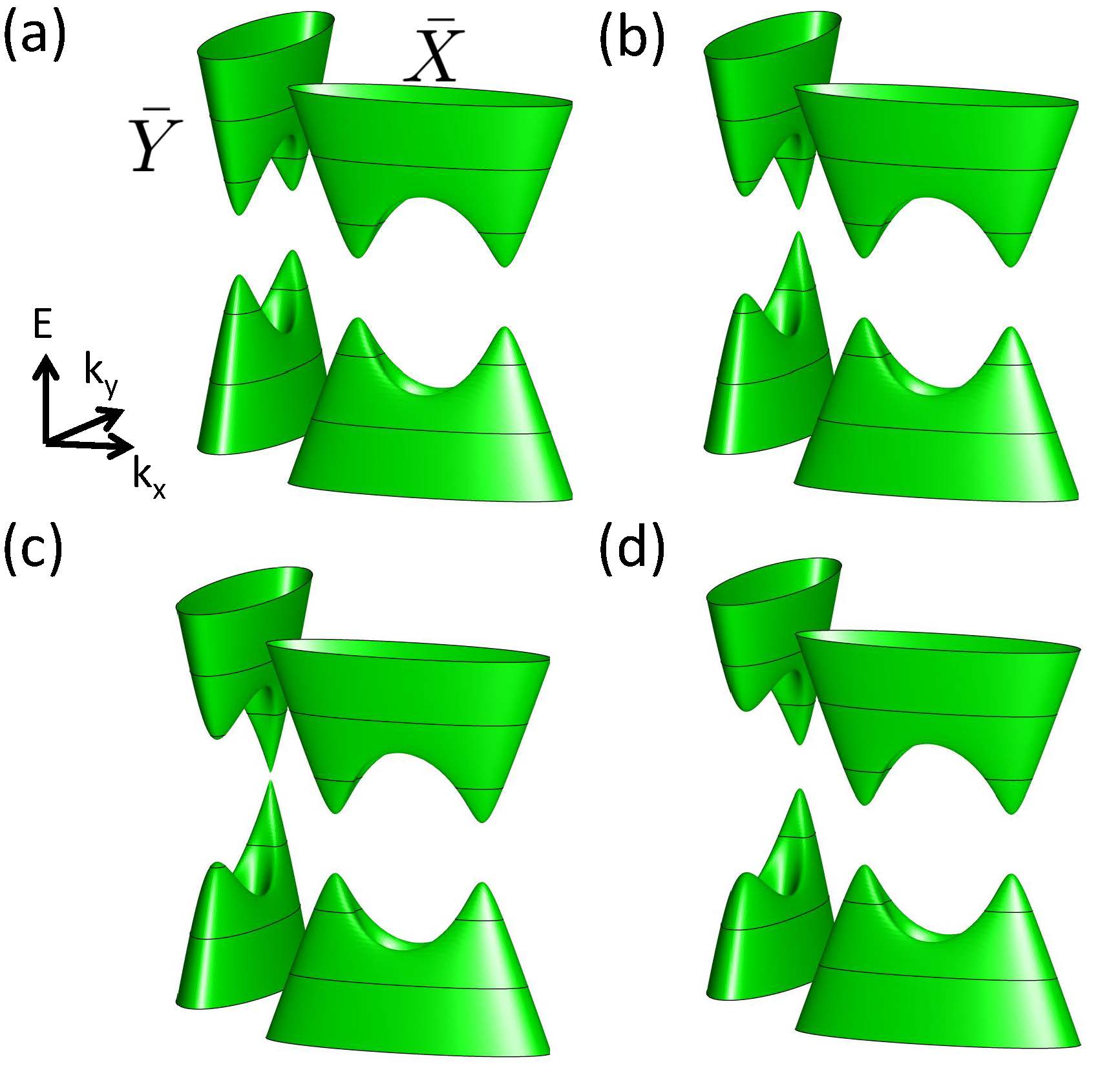}
\caption{(Color online) Evolution of band gaps in the Dirac cones under strain perturbation $V_s\Sigma_{23}$ with a fixed Zeeman field, where $h_z=0.1$. The strength of the strain perturbation is given by $V_s=$ (a) 0, (b) 0.03, (c) 0.05, and (d) 0.08. From (b) to (d), one of the Dirac cones around $\bar{Y}$ closes and reopens again due to a band inversion. The model parameters are the same as those used in Fig.~\ref{fig:b1}.}
\label{fig:biv}
\end{center}
\end{figure}

%(key point is to emphasize strain can induce Zeeman-like effect, without %breaking TRS)

\subsection{Tuning the Chern number via the interplay of strain and a Zeeman fields}
Like a perpendicular Zeeman field, the strain field on the surface of a TCI may not only give rise to spin reorientation, but it may also induce a charge gap in the Dirac cones. However, there are key differences in the effects of strain and Zeeman fields. An applied Zeeman field, which respects $C_2$ rotational symmetry around the $z$-axis, opens gaps with the {\it same} sign in the pair of Dirac cones around $\bar{X}$, while an applied strain field, which breaks $C_2$ rotational symmetry, would induce gaps with {\it opposite} sign. This observation provides the foundation for tenability of the Chern number via the interplay of applied strain and Zeeman fields, suggested in Ref.\onlinecite{Fang14} using a simplified two-band model for each Dirac point.

As a concrete example based on our four-band model, consider an applied Zeeman field along the $z$ direction along with a strain field which breaks $M_{yz}$ but preserves $M_{xz}$ around $\bar{Y}$ on the (001) surface. From symmetry considerations, the strain and Zeeman fields around $\bar{X}$ and $\bar{Y}$ become coupled as 
\ba
H_{\bar{X}} &+& h_z\Sigma_{03}, \nonumber \\
H_{\bar{Y}} &+& h_z\Sigma_{03}+V_s\Sigma_{23},
\ea
where $h_z$ ($V_s$) is the field strength of Zeeman (strain) field. Note that for a given $h_z$, all four low-energy Dirac points open gaps in the absence of an applied strain [see Fig.~\ref{fig:biv}(a)]. When a non-vanishing $V_s\ll \frac{1}{2}h_z$ is introduced, the gaps of the two massive Dirac cones around $\bar{Y}$ evolve in an opposite manner in the sense that one cone increases while the other decreases. This is consistent with a picture in which the strain field induces out-of-the-plane spin components around the two massive Dirac cones which form time-reversed partners with respect to their spin textures [see Fig.~\ref{fig:f1}(b)], and therefore, respond oppositely to the existing $h_z$ field. With increasing strength of $V_s$, the gap in the Dirac cone with a decreasing gap continues to decrease further, becomes gapless at $V_s = \frac{1}{2}h_z$, and reopens inverted again, as shown in Figs.~\ref{fig:biv}(b)-(d). This band inversion indicates that the system undergoes a topological phase transition.

\begin{figure}[tbh]
\begin{center}
\includegraphics[width=200pt]{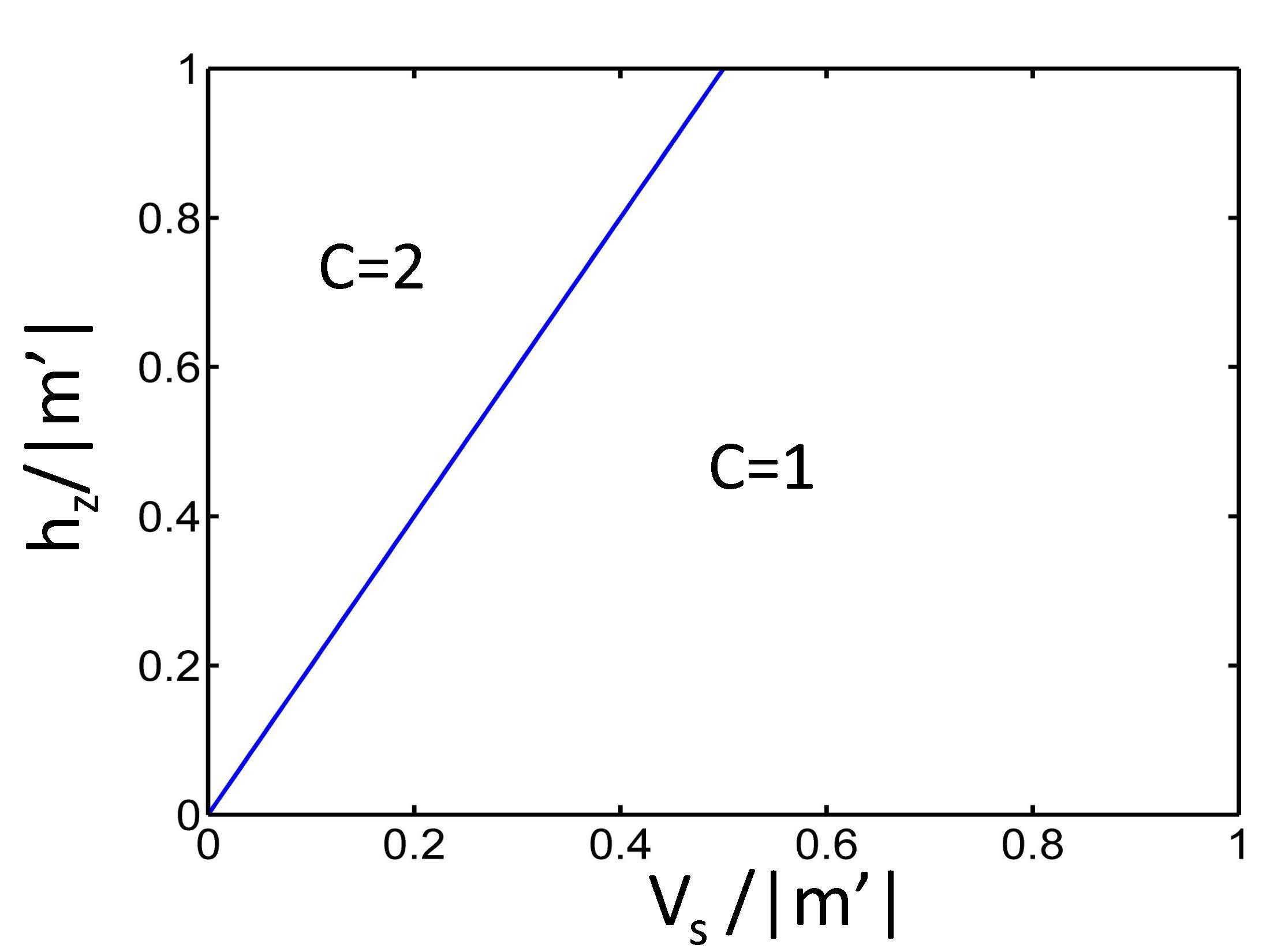}
\caption{(Color online) Chern number for our four-band model of a TCI as a function of $V_s$ and $h_z$. Model parameters are fixed as: $m$=-2, $m'$=-1, $v_{1x}$=-1, $v_{1y}$=-1.5, with other parameters taken to be zero. The phase boundary (blue line) is given by $h_z=2V_s$.}
\label{fig:Chern}
\end{center}
\end{figure}

The topological nature of a ground state with broken time-reversal symmetry can be characterized by the Chern number. A non-zero value of the Chern number, $C$, given by
\begin{align}
C=\frac{-1}{\pi }\sum_{n\in\text{occ.}}\sum_{m\neq n}\frac{\text{Im}[\langle u_n|\frac{\partial H}{\partial k_x}|u_m\rangle\langle u_m|\frac{\partial H}{\partial k_y}|u_n\rangle]}{(E_n-E_m)^2},
\end{align}
where the summation over momentum space is implicit, indicates a finite Hall conductance $\sigma=C\frac{e^2}{h}$, which can be obtained by integrating the Berry curvature of the wave functions of the occupied bands over the momentum space\cite{Andrei13}. Fig.~\ref{fig:Chern} presents the Chern number of the system as a function of the Zeeman field strength $h_z$ and the strain strength $V_s$ within the framework of our four-band model. In the presence of a Zeeman field, by varying the strength of strain from small to large values is seen to drive the system from $C=2$ to $C=1$, demonstrating the tunability of the system to the strain field. Note that in our consideration here we have implicitly assumed that the sample is thick enough so that the hybridization between the top and bottom surfaces of the sample can be neglected; with $h_z\neq 0$ the bottom surface will contribute another Chern number $C=2$, which is not shown in Fig.~\ref{fig:Chern} for simplicity.

\section{EFFECTS OF ELECTRON-ELECTRON INTERACTION}
%Consider short-range Coulomb repulsion; particle-hole channels only; mean-field %calculation; competing orders and phase diagram

%Wei-Feng and Hsin, I suggest removing the entire following paragraph since it is somewhat repetitive, and does not add that much beyond what has been discussed already in the manuscript earlier. I have not revised this paragraph for this reason. I will come back to it, if you feel strongly that this should not be removed. ARUN

%Wei-Feng and Hsin: Please note that some references might get screwed up if the following paragraph is removed. ARUN

%A system with multiple Dirac-like electronic structures, such as graphene or a weak topological insulator, can usually host a number of competing orders due to non-negligible electron-electron interactions whose presence may be justified by the relatively poor charge screening effect\cite{CastroNeto09,Coulomb}. The low-energy electronic structure in a TCI as we focused here is, in fact, very similar to that of a weak TI: Both of them have even number of surface Dirac cones with non-trivial spin texture due to strong spin-orbit coupling. There are, however, a few significant differences. Firstly, they have different spin and orbital texture\cite{Yan12,Zeljkovic14}. Secondly, the gapless Dirac cones in a TCI are guaranteed by the mirror symmetries, instead of the TRS. As a result, the location of each Dirac point is on a mirror line and not at a time-reversal invariant momentum. Therefore, one may anticipate to see a distinct phase diagram for a TCI under electron-electron interactions. 

In discussing electron-electron interactions, we consider short-range repulsive interactions between electrons on the (001) surface of a TCI, and focus on delineating how the phase diagram evolves in the presence of a strain, which breaks one of the mirror symmetries, say, $M_{yz}$, based on the the interacting Hamiltonian: 
\begin{equation}
\hat{H}=\sum_{\bk}\hat{\psi}^{\dagger}(\bk)H_{(001)}\hat{\psi}(\bk)+\hat{H}_{int},
\end{equation}
where
\ba
\hat{\psi}^{\dagger}(\bk)&=&[c^{\dagger}_{\bar X,p_z\uparrow}(\bk),c^{\dagger}_{\bar X,p_z\downarrow}(\bk),
c^{\dagger}_{\bar X,p_x\uparrow}(\bk),c^{\dagger}_{\bar X,p_x\downarrow}(\bk),\nonumber \\
&& c^{\dagger}_{\bar Y,p_z\uparrow}(\bk),c^{\dagger}_{\bar Y,p_z\downarrow}(\bk),
c^{\dagger}_{\bar Y,p_y\uparrow}(\bk),c^{\dagger}_{\bar Y,p_y\downarrow}(\bk)],
\label{eq:basis}
\ea
with the subscript $\bar X$ ($\bar Y$) denoting the momentum point involved in the $\bk$ expansion. For $H_{(001)}$, we have added the strain perturbation and hence, $H_{(001)}=H_{\bar{X}}\oplus (H_{\bar{Y}}+V_s\Sigma_{23})$. The interaction term, $\hat{H}_{int}=\hat{H}_{U}+\hat{H}_{V}$; $\hat{H}_{U}$ and $\hat{H}_{V}$ denote repulsive contact interactions between like and unlike orbitals, respectively, with
\ba
\hat{H}_{U} &=& U\int d^2 r\;\sum_{\eta=p_z,p_x,p_y}\hat{n}_{\eta\uparrow}({\br})
\hat{n}_{\eta\downarrow}({\br}), \nonumber \\
\hat{H}_{V} &=& \frac{V}{2}\int d^2 r\;\sum_{\eta\neq\eta^\prime}\hat{n}_{\eta}({\br})
\hat{n}_{\eta^\prime}({\br}),
\label{Hint}
\ea
where $\hat{n}_{\eta}(\br)=\sum_{s}\hat{n}_{\eta s}(\br)=\sum_{\mathbf{Q},s}\tilde{c}^\dagger_{\mathbf{Q},\eta s}\tilde{c}_{\mathbf{Q},\eta s}(\br)$ ($\mathbf{Q}=\bar{X},\bar{Y}$, $\eta=p_x,p_y,p_z$) [$\tilde{c}_{\mathbf{Q},\eta s}(\br)$ are defined in Eq.~(\ref{eq:FTc}) below].

Since we are treating only the low-energy, long-wavelength physics, it is reasonable to consider field operators $c_{\mathbf{Q},\eta\sigma}(\br)$, which vary slowly on the scale of the lattice constant. This can be done by Fourier transforming the operators:
\be
\tilde{c}_{\mathbf{Q},\eta s}(\br) = \sum_{\mathbf{K}}e^{i\mathbf{K}\cdot\br}
c_{\mathbf{Q},\eta s}(\mathbf{K}) = e^{i\mathbf{Q}\cdot\br}
c_{\mathbf{Q},\eta s}(\br),
\label{eq:FTc}
\ee
where $\mathbf{K}=\mathbf{Q}+\bk$, expanding from the origin (0,0). In terms of these field and density operators, and the identities given in the Appendix, the full Hamiltonian $\hat{H}$ can be rewritten as
\begin{equation}
\hat{H}=\int d^2 r\;\hat{\Psi}^{\dag}H_{(001)}\hat{\Psi}+\hat{H}_{int},
\end{equation}
where $\hat{\Psi}^{\dagger}(\br)$ has the same form as Eq.~(\ref{eq:basis}) except that the $\bk$-dependence is now replaced by $\br$. 

%density operator...

In order to determine the ground state of the system as a function of the interaction strength and fermion density, we employ the self-consistent mean-field (MF) approach, which is expected to be reliable as long as the interaction strength is much smaller than the bulk band gap. Taking $U=V$ as a representative case, we decouple the $\hat{H}_{int}$ into bilinear fermion terms. After some straightforward but tedious algebra, we obtain 52 order parameters as well as 12 renormalized band parameters as detailed in the Appendix, which can be generally expressed as
\be
O_{\gamma\alpha\beta}=\langle\hat{\Psi}^{\dagger}\Sigma_{\gamma\alpha\beta}
\hat{\Psi}\rangle
%=\int_{k<k_{\Lambda}}^{E=\mu}\frac{d^2k}{(2\pi)^2}\langle
%\Sigma_{h\alpha\beta}\rangle,
\label{eq:order}
\ee
where a summation over momentum and occupied states is implicit, and $\Sigma_{\gamma\alpha\beta}=\tau_{\gamma}\otimes\Sigma_{\alpha\beta}$, with the Pauli matrices, $\tau_{\gamma}$, acting on $\bar{X}\bar{Y}$ pseudo-spin (valley) space. Note that, here we consider only the symmetry breaking orders in the particle-hole channel: superconductivity in the particle-particle channel due to weak onsite repulsion can only be achieved via beyond MF treatment, usually with exponentially small transition temperature\cite{Kohn65}. 

%(except at van Hove singularity [**]).  

In this connection, we define the ``particle density'' at a given chemical potential $\mu$ as the density deviation from the total particle density where $\mu=E_d=0$, namely,
\be
n=\int_{k<k_{\Lambda}}^{E(\bk)=\mu} \frac{d^2k}{(2\pi)^2}-\int_{k<k_{\Lambda}}^{E_0(\bk)=0} \frac{d^2k}{(2\pi)^2},
\label{eq:density}
\ee
where $E_0(\bk)$ ($E(\bk)$) is the eigenenergy of the non-interacting (MF-decoupled interacting) system; $k_{\Lambda}$ is a cutoff momentum, which is chosen such that our results are insensitive to its magnitude. Note that by any induced valley ``polarization'', $O_{300}$, we mean
\begin{equation}
O_{300}=\langle\hat{\Psi}^{\dagger}\Sigma_{300}
\hat{\Psi}\rangle
-\int^{E_0(\bk)=0}_{k<k_{\Lambda}}\frac{d^2k}{(2\pi)^2}\langle\Sigma_{300}\rangle.
\label{eq:orbital}
\end{equation}
That is, the ``valley polarization'' due to external strain in the absence of interactions is subtracted in defining $O_{300}$.

\begin{figure}
\begin{center}
\includegraphics[width=200pt]{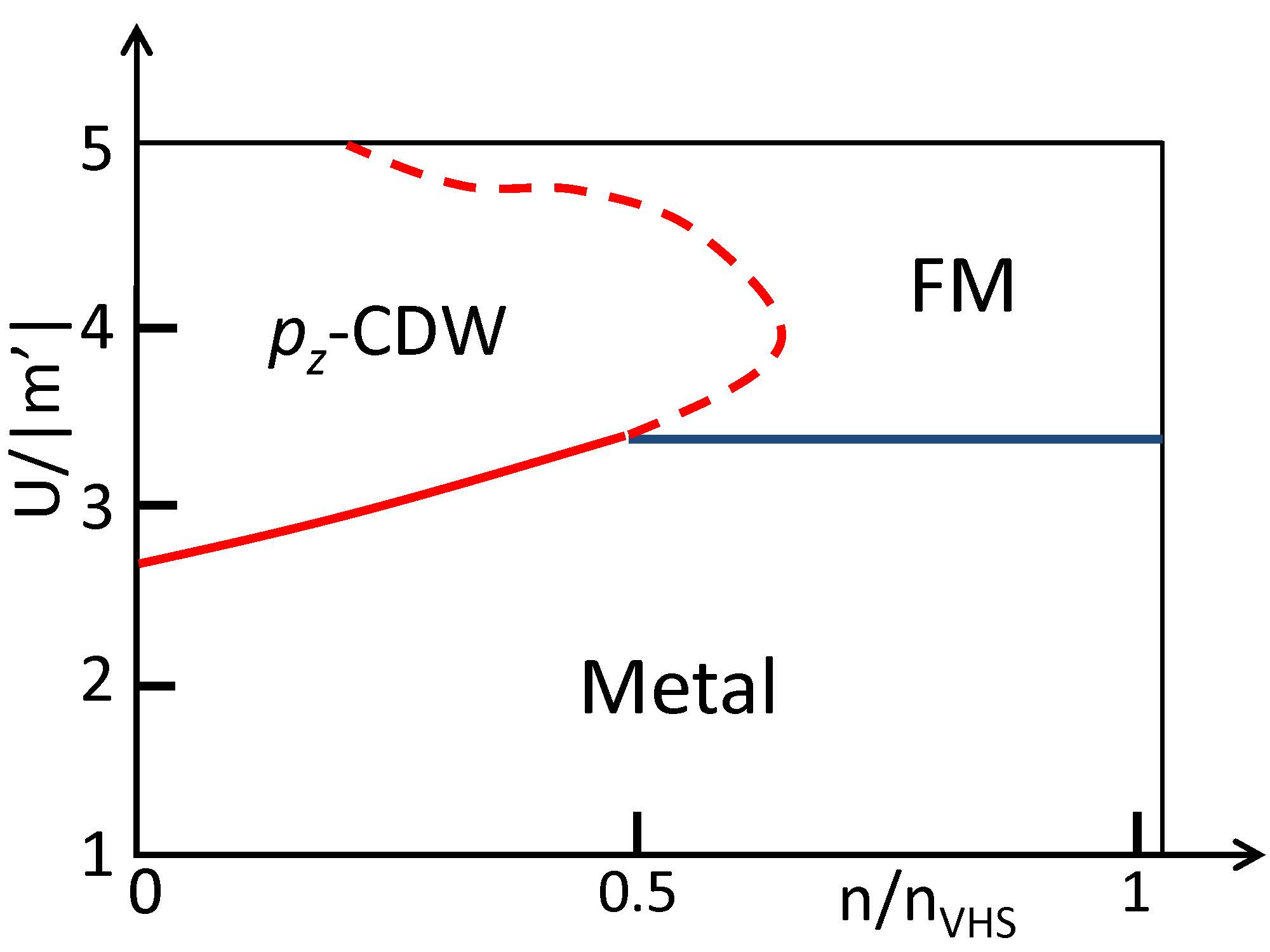}
\caption{(Color online) $U$ versus $n$ phase diagram without strain. $n_{\text{VHS}}$ denotes the particle density at $E_{VHS+}$. Solid and dashed lines mark phase boundaries of continuous and abrupt phase transitions, respectively.}
\label{fig:PD1}
\end{center}
\end{figure}

For a given $U$ and $n$, the ground state can then be determined by minimizing the MF free energy with respect to various order parameters using the form detailed in the Appendix. We numerically solve the resulting set of coupled equations self-consistently to obtain the zero-temperature $U$-$n$ phase diagrams both with and without the applied strain. Note that, in order to make the underlying physics more transparent, we use the following model parameters in computations: $m=-2$, $m'=-1$, $v_{1x}=-1$, $v_{1y}=-1.5$, and a cutoff energy, $\Lambda\equiv E_{0}(k_{\Lambda})=2\sqrt{m^2+m'{}^2}$, which determines $k_{\Lambda}$. However, we expect our results to be generic and relevant more generally, detailed effects of material-specific parameters notwithstanding.

\subsection{System without strain}
We begin by considering a system without strain, {\it i.e.}, $V_s=0$. In the order parameter space we explored in this case, the leading orders (aside from the metal phase) are basically the ferromagnetic state (FM), $O_{003}$, and the $p_z$-orbital rich $(\pi,\pi)$ charge density wave ($p_z$-CDW): 
\be
\frac{O_{100}+O_{130}}{2}=\Big\langle\sum_{\br}(-1)^{x+y}[\hat n_{p_z,\uparrow}(\br)+\hat n_{p_z,\downarrow}(\br)]\Big\rangle, 
\ee
where the length scale is in units of the lattice constant.  The zero-temperature $U$-$n$ phase diagram is shown in Fig.~\ref{fig:PD1}. The $p_z$-CDW phase corresponds to a non-vanishing $\frac{O_{100}+O_{130}}{2}$ (the leading piece), and the FM phase represents a non-vanishing value of $O_{003}$ (the leading piece). Finally, in the metal phase, all symmetries are preserved with vanishing values of all 52 order parameters.

\begin{figure}[th]
\begin{center}
\includegraphics[width=8cm]{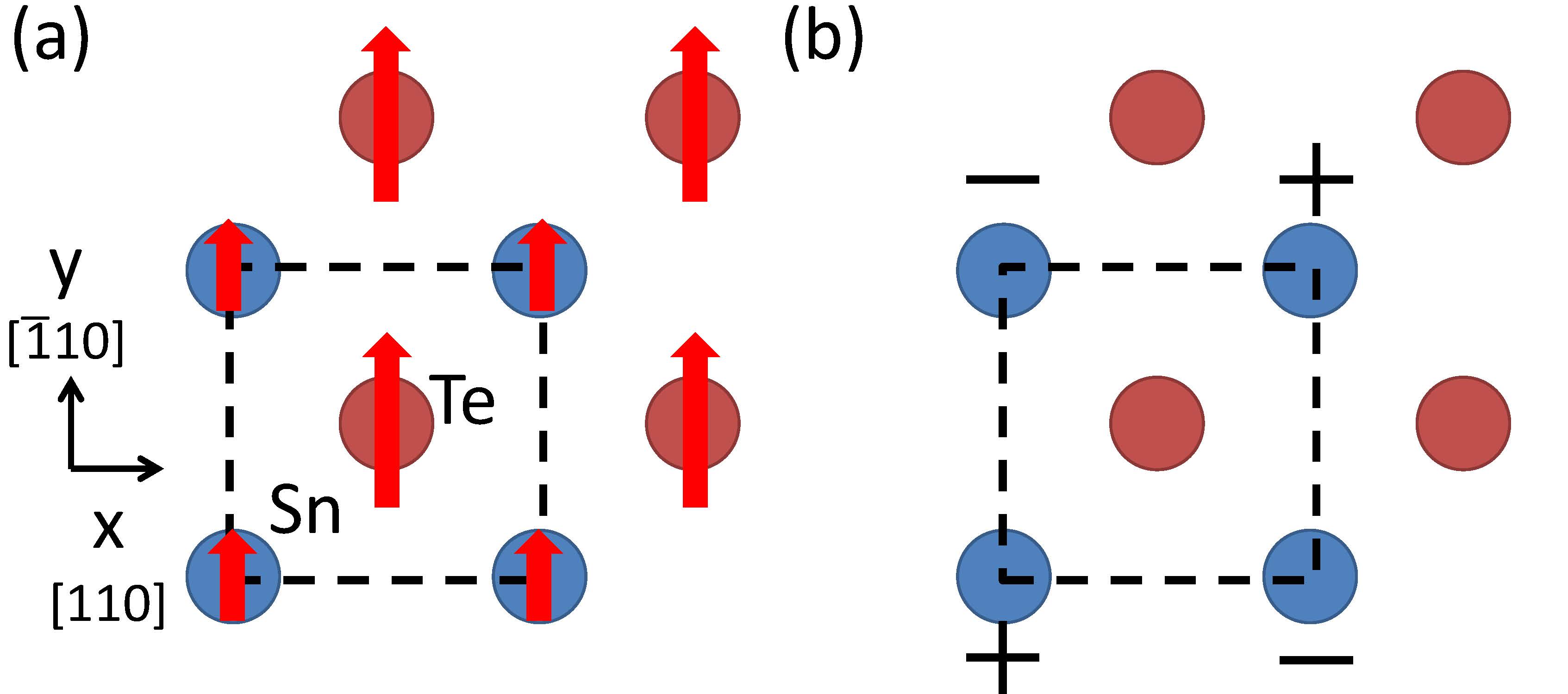}
\caption{(Color online) Schematic plots illustrating (a) the ferromagnetic state (FM), and (b) the $p_z$-orbital rich, $(\pi,\pi)$ charge density wave state ($p_z$-CDW). Different colors label the two distinct atoms involved; arrows depict the net out-of-the-plane spin polarization.}
\label{fig:realspace}
\end{center}
\end{figure}

At $n=0$, the ``$p_z$-CDW'' phase appears when the interaction strength $U$ is larger than the critical value $U_c\approx 2.7$. This phase involves a finite value of $\frac{O_{100}+O_{130}}{2}$; it is associated with broken translation symmetry, while the TRS and both the mirror symmetries remain intact [see schematic Fig.~\ref{fig:realspace}(b)], so that each Dirac cone remains gapless. The existence of a critical value of $U$ can be understood by noticing that in the non-interacting limit, there are only four Dirac points with zero density of states at the Fermi level. The necessity of a threshold value of $U$ has also been predicted theoretically in 2D systems with linear energy dispersion, such as graphene and the surfaces of 3D weak TIs\cite{Raghu08,Liu2012906}.

Upon electron doping, the system assumes either the FM phase [see Fig.~\ref{fig:realspace}(a)] at relatively high particle densities, or the $p_z$-CDW phase [see Fig.~\ref{fig:realspace}(b)] at low particle densities. The FM phase is associated with broken TRS as well as both the mirror symmetries simultaneously, resulting in gaps to open up at all the Dirac points, although the corresponding ground state is still metallic with finite $\mu$. Note that the critical $U$, above which the system enters into the gapless $p_z$-CDW phase, is less than the critical value for the FM phase, indicating a favorable free energy gain compared to the competing FM phase at low doping.

%Wei-Feng and Hsin: Please review the following paragraph specially carefully, as I was not sure what was meant in some parts - please note the suggestion for an added footnote in this pargarph--ARUN

We further note several points in connection with the phase diagram of Fig.~\ref{fig:PD1} as follows: (1) The spin density wave (SDW) phase, as anticipated usually in Dirac systems like graphene, does not occur here at $n=0$ for large $U$. This could be attributed to the non-trivial orbital textures and strong spin-orbit coupling in our case; (2) We find that the dynamically generated strain-like order, which could gap the spectrum, is relatively disfavored as the system tries to preserve the combined $C_{2T}$ symmetry, although such a phase becomes favored in the presence of external strain; (3) If we suppress inter-valley scattering, our preliminary results suggest that the CDW phase is also suppressed\cite{long-range}; and finally, (4) As we pointed out already in Sec. II above, the density of states at an energy scale $\omega$ near a van Hove singularity diverges like $\ln |\omega|$, indicating propensity for the occurrence of more symmetry breaking orders (including those in the particle-particle channel). The present mean-field treatment would then be inappropriate, even in the weak-coupling limit.

\subsection{System under moderate strain}
\begin{figure}
\begin{center}
\includegraphics[width=200pt]{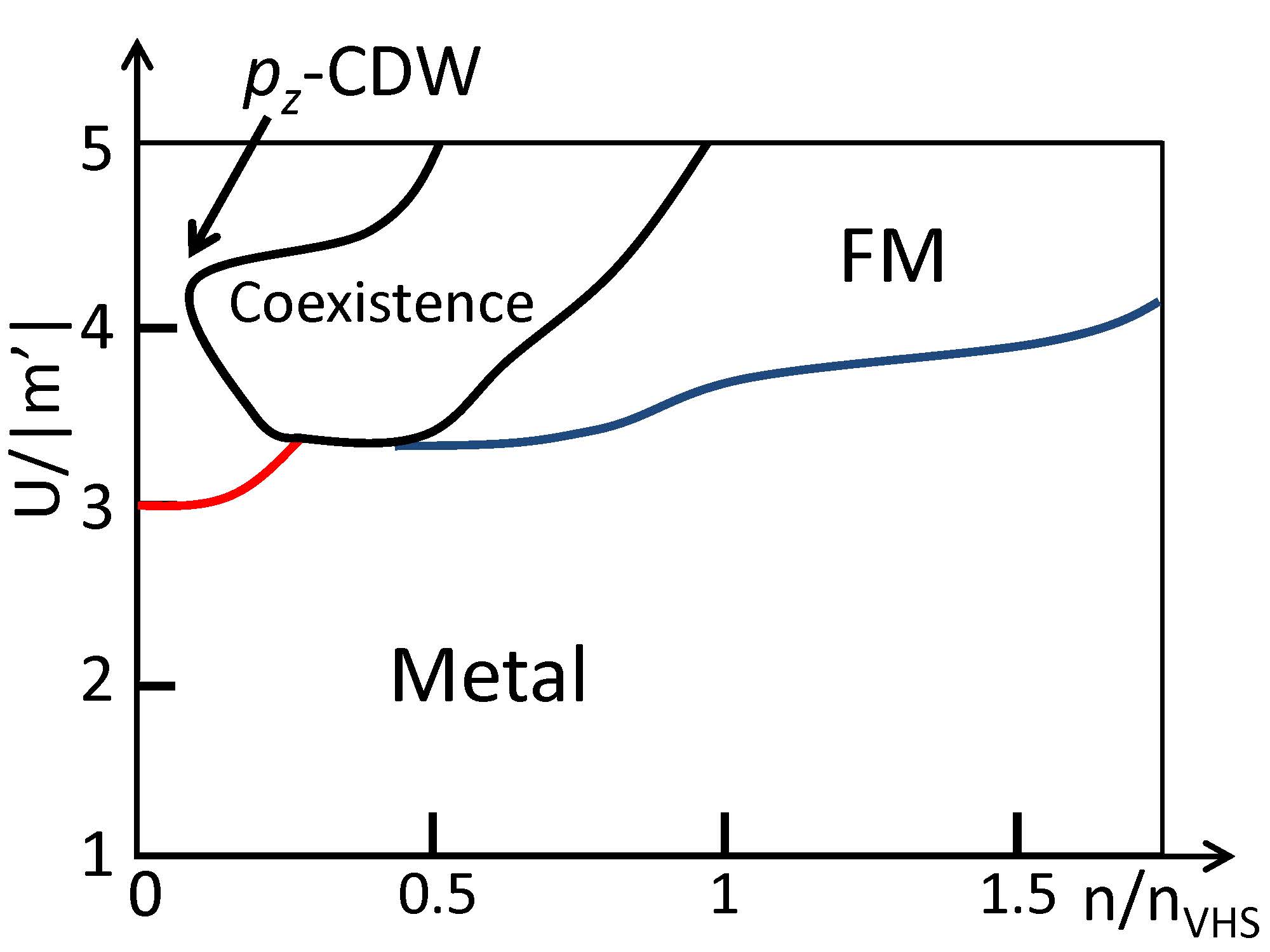}
\caption{(Color online) $U$ versus $n$ phase diagram in the presence of a moderate strain. Lines mark boundaries between various continuous phases.}
\label{fig:PD2} 
\end{center}
\end{figure}

Here we consider an applied strain with a moderate coupling strength, $V_s=|m'|$ [of order $\mathcal{O}(E_{VHS+}$)], which breaks the $yz$-mirror symmetry. The resulting zero-temperature phase diagram is shown in Fig.~\ref{fig:PD2}. In contrast to the case without strain, a coexisting phase lying between the $p_z$-CDW and FM phases is now seen to emerge. Suppression of the $p_z$-CDW phase with strain can be anticipated on physical grounds. The reason is that any $(\pi,\pi)$ order requires good nesting between the valleys, but the strain weakens the nesting tendency by opening gaps at Dirac points along the $yz$-mirror line. Notably, in the strained system, spin polarization in the FM phase is mainly contributed by the $p_x$ orbital due to the broken mirror symmetry, which lifts the $p_y$ orbital to higher energies.

\subsection{System with large strain}
\begin{figure}
\begin{center}
\includegraphics[width=200pt]{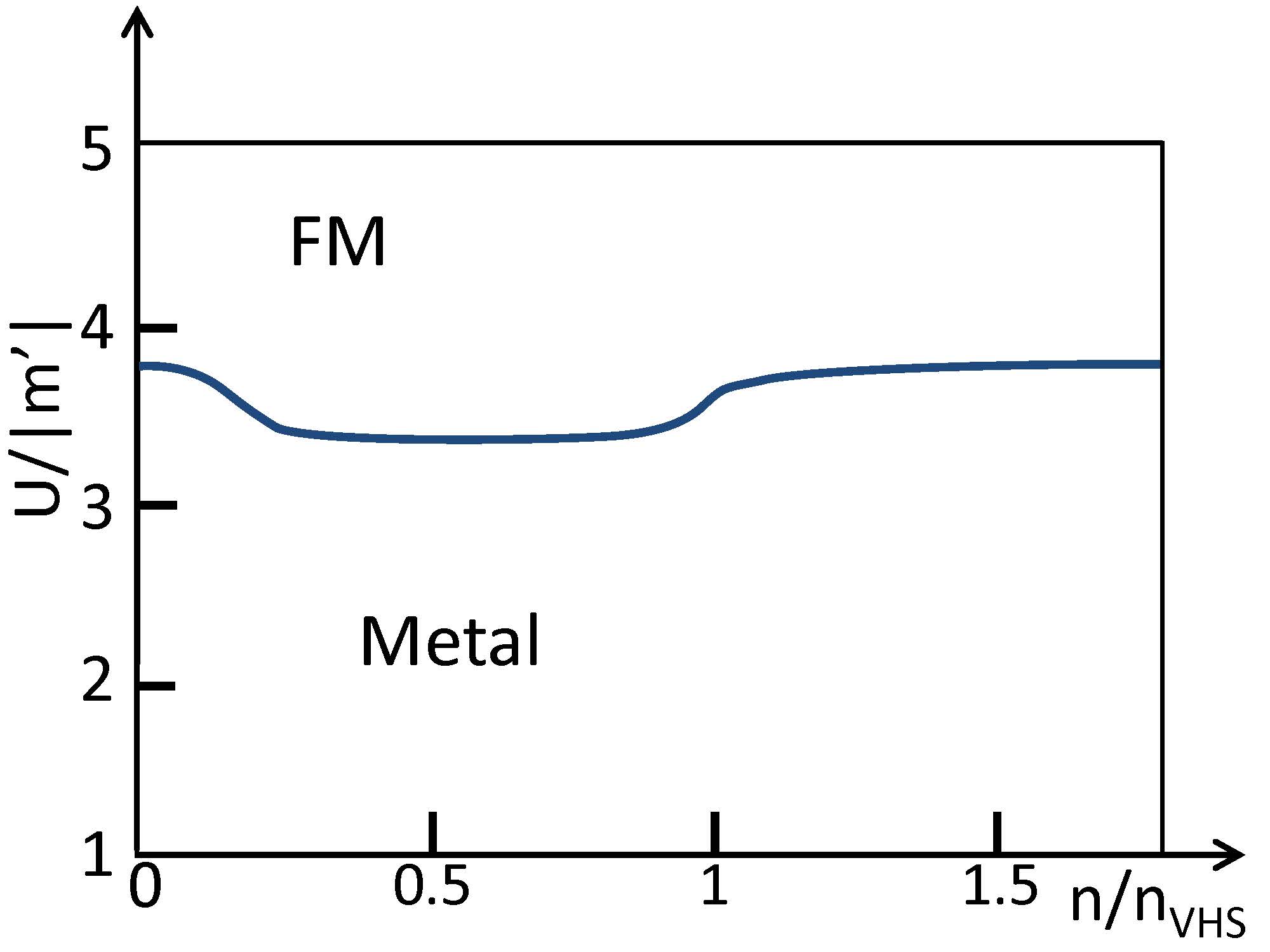}
\caption{(Color online) $U$ versus $n$ phase diagram in the case of large strain. Solid line marks the boundary between the two continuous phases.}
\label{fig:PD3}
\end{center}
\end{figure}

When the coupling strength of the applied strain $V_s$ becomes the largest energy scale in the system, all other degrees of freedom can be ignored at the $\bar Y$ valley as states around $\bar Y$ are all gapped out, leading to the phase diagram of Fig.~\ref{fig:PD3}. All $(\pi,\pi)$ orders are seen to disappear, and only the FM phase survives for $U>U_c(n)$ at a given density $n$. The presence of the concave boundary region separating the metallic and FM phases in Fig. 8 can be understood from energetics: At any given density in this region, the chemical potential lies within the gap induced by the FM order at $E_{d2}$, yielding a gain in free energy and thus lowers $U_c$. As the particle density increases, the chemical potential is eventually unable to locate within the gap and this argument fails. Interestingly, at $n=0$, although the ground state is insulating with a small gap due to TRS breaking, the system does not exhibit quantum anomalous Hall effect. This should be contrasted sharply with the case of a Zeeman field applied perpendicular to the (001) surface of a TCI in the non-interacting limit, where such a broken TRS leads to non-vanishing Chern number as shown in Sec. IIIB above.

\section{Discussion and conclusion}
%Phase transition by tuning strain (application potential); analogy with graphene

Some of the strains discussed in this study could be realized experimentally through chemical doping. Since PbSe and SnSe assume cubic and orthorhombic structures\cite{structure}, respectively, strains or lattice distortions could likely be manipulated by adjusting chemical compositions. In fact, a ferroelectric-like lattice distortion has been observed in the TCIs via STM/STS spectroscopy\cite{OkadaTCIstrain}. Mechanical tuning could provide another approach. For example, a piezoelectric layer could be grown on the surface of a TCI to allow manipulation of strains via electric field as proposed in Ref.\onlinecite{Fang14}.

As to practical applications, we have shown clearly that the Chern number of the surface states can be tuned through the interplay of applied strain and Zeeman fields. This provides a potential new handle for controlling the topological conducting edge channels for low power-consuming, next generation electronic technologies. Furthermore, our analysis shows that in the presence of electron-electron interactions, ferromagnetism could be turned on/off via strain fields, which, offering a novel pathway toward spintronics applications.

In summary, we have systematically investigated the effects of strains, which break either one or both mirror symmetries without breaking time-reversal symmetry, on the (001) surface of the TCIs. Under mirror-symmetry breaking strains, the low energy Dirac cones not only become massive, but also develop hedgehog-like spin textures. We show that the Chern number of the surface states can be tuned via applied strain fields in the presence of a Zeeman field. Finally, we delineate the competing orders that can result from varying strengths of short-range (repulsive) electron-electron interactions at the mean-field level, and show that the resulting correlated phases are amenable to control through strain fields. The tunability of the interacting as well as the non-interacting electronic structures and topological states of the TCIs via strain fields and their interplay with Zeeman fields as revealed by our study suggests new pathways for developing spintronics and other applications platforms based on the TCIs.

\begin{acknowledgments}
We thank Chen Fang and Hong Yao for useful discussions. W.F.T. and C.Y.H. acknowledge the support from MOST in Taiwan under Grant No.103-2112-M-110-008-MY3. The work at Northeastern University was supported by the US Department of Energy (DOE), Office of Science, Basic Energy Sciences grant number DE-FG02-07ER46352, and benefited from Northeastern University's Advanced Scientific Computation Center (ASCC) and the NERSC supercomputing center through DOE grant number DE-AC02-05CH11231. H.L. acknowledges the Singapore National Research Foundation for the support under NRF Award No. NRF-NRFF2013-03.
\end{acknowledgments}

\appendix
\section{\label{A1} Derivation of mean-field theory in Sec. IV}
Here we present further details of the MF theory used in Sec. IV. In terms of the slowly varying field operators defined in Eq.~(\ref{eq:FTc}), the relevant number operators are given by
\ba
\hat{n}_{p_z\uparrow}({\bf r})&=&c_{\bar X,p_z\uparrow}^{\dag}c_{\bar X,p_z\uparrow}+c_{\bar Y,p_z\uparrow}^{\dag}c_{\bar Y,p_z\uparrow}\notag\\
&+&(-1)^{x+y}(c_{\bar X,p_z\uparrow}^{\dag}c_{\bar Y,p_z\uparrow}+c_{\bar Y,p_z\uparrow}^{\dag}c_{\bar X,p_z\uparrow}),\notag\\
\hat{n}_{p_z\downarrow}({\bf r})&=&c_{\bar X,p_z\downarrow}^{\dag}c_{\bar X,p_z\downarrow}+c_{\bar Y,p_z\downarrow}^{\dag}c_{\bar Y,p_z\downarrow}\notag\\
&+&(-1)^{x+y}(c_{\bar X,p_z\downarrow}^{\dag}c_{\bar Y,p_z\downarrow}+c_{\bar Y,p_z\downarrow}^{\dag}c_{\bar X,p_z\downarrow}),\notag\\
\hat{n}_{p_x\uparrow}({\bf r})&=&c_{\bar X,p_x\uparrow}^{\dag}({\bf r})c_{\bar X,p_x\uparrow}({\bf r}),\notag\\
\hat{n}_{p_x\downarrow}({\bf r})&=&c_{\bar X,p_x\downarrow}^{\dag}({\bf r})c_{\bar X,p_x\downarrow}({\bf r}),\notag\\
\hat{n}_{p_y\uparrow}({\bf r})&=&c_{\bar Y,p_y\uparrow}^{\dag}({\bf r})c_{\bar Y,p_y\uparrow}({\bf r}),\notag\\
\hat{n}_{p_y\downarrow}({\bf r})&=&c_{\bar Y,p_y\downarrow}^{\dag}({\bf r})c_{\bar Y,p_y\downarrow}({\bf r}),\notag\\
\hat{n}_{p_z}({\bf r})&=&\hat{n}_{p_z\uparrow}({\bf r})+\hat{n}_{p_z\downarrow}({\bf r}),\notag\\
\hat{n}_{p_x}({\bf r})&=&\hat{n}_{p_x\uparrow}({\bf r})+\hat{n}_{p_x\downarrow}({\bf r}),\notag\\
\hat{n}_{p_y}({\bf r})&=&\hat{n}_{p_y\uparrow}({\bf r})+\hat{n}_{p_y\downarrow}({\bf r}).
\ea
%According to Eq.~(\ref{Hint}), the summation in interaction as $U=V$ is given by
%\begin{align}
%&U\sum_{\eta=p_z,p_x,p_y}\hat{n}_{\eta\uparrow}({\bf r})\hat{n}_{\eta\downarrow}({\bf r})
%+\frac{U}{2}\sum_{\eta\neq\eta'}\hat{n}_{\eta}({\bf r})\hat{n}_{\eta'}({\bf r})\notag\\
%=&U\big[\hat{n}_{p_z,\uparrow}({\bf r})\hat{n}_{p_z,\downarrow}({\bf r})
%+\hat{n}_{p_x,\uparrow}({\bf r})\hat{n}_{p_x,\downarrow}({\bf r})
%+\hat{n}_{p_y,\uparrow}({\bf r})\hat{n}_{p_y,\downarrow}({\bf r})\notag\\
%+&\big(\hat{n}_{p_z,\uparrow}({\bf r})+\hat{n}_{p_z,\downarrow}({\bf r})\big)
%\big(\hat{n}_{p_x,\uparrow}({\bf r})+\hat{n}_{p_x,\downarrow}({\bf r})\big)\notag\\
%+&\big(\hat{n}_{p_z,\uparrow}({\bf r})+\hat{n}_{p_z,\downarrow}({\bf r})\big)
%\big(\hat{n}_{p_y,\uparrow}({\bf r})+\hat{n}_{p_y,\downarrow}({\bf r})\big)\notag\\
%+&\big(\hat{n}_{p_x,\uparrow}({\bf r})+\hat{n}_{p_x,\downarrow}({\bf r})\big)
%\big(\hat{n}_{p_y,\uparrow}({\bf r})+\hat{n}_{p_y,\downarrow}({\bf r})\big)\big].\label{ne1}
%\end{align}

We illustrate our MF treatment with the example of $\int d^2r\;U\hat{n}_{p_x,\uparrow}({\bf r})\hat{n}_{p_x,\downarrow}({\bf r})$. For this purpose, we decouple bilinear fermion operators composed of $\hat A$ or $\hat B$ in the particle-hole channel into $\hat A\langle \hat B\rangle+\hat B\langle \hat A\rangle-\langle\hat B\rangle\langle \hat A\rangle + \cdots$, where $\cdots$ represents fluctuations away from the mean-field values, which are neglected. The condensation energy, $-\langle\hat B\rangle\langle \hat A\rangle$, for $\int d^2r\;U\hat{n}_{p_x,\uparrow}({\bf r})\hat{n}_{p_x,\downarrow}({\bf r})$ is thus given by
\begin{align}
&-U\Big\langle c_{\bar X,p_x\uparrow}^{\dag}c_{\bar X,p_x\uparrow}\Big\rangle
\Big\langle c_{\bar X,p_x\downarrow}^{\dag}c_{\bar X,p_x\downarrow}\Big\rangle\notag\\
&+U\Big\langle c_{\bar X,p_x\uparrow}^{\dag}c_{\bar X,p_x\downarrow}\Big\rangle
\Big\langle c_{\bar X,p_x\downarrow}^{\dag}c_{\bar X,p_x\uparrow}\Big\rangle,\label{demo1}
\end{align}
where the ${\bf r}$-dependence is implicit. In fact,
$\Big\langle c_{\bar X,p_x\uparrow}^{\dag}c_{\bar X,p_x\uparrow}\Big\rangle$ can be further rewritten as
\begin{align}
&\frac{S}{8}\Big\langle\hat{\Psi}^{\dag}(\Sigma_{000}+\Sigma_{003}-\Sigma_{030}-\Sigma_{033}\notag\\
&+\Sigma_{300}+\Sigma_{303}-\Sigma_{330}-\Sigma_{333})\hat{\Psi}\Big\rangle\notag\\
&=\frac{S}{8}(O_{000}+O_{003}-O_{030}-O_{033}\notag\\
&+O_{300}+O_{303}-O_{330}-O_{333}),
\label{demo2}
\end{align}
where $O_{\gamma\alpha\beta}$ was defined in Eq.~(\ref{eq:order}) and $S$ is a unit area.
Notice that $O_{000}$ is the particle density defined in Eq.~(\ref{eq:density}). When the similar procedure is applied to other terms in Eq.~(\ref{demo1}), it turns out that $-\langle\hat B\rangle\langle \hat A\rangle$ term becomes:
\begin{align}
-S\frac{U}{64}&[(O_{000}-O_{030}+O_{300}-O_{330})^2\notag\\
&-(O_{003}-O_{033}+O_{303}-O_{333})^2]\notag\\
+S\frac{U}{64}&[(O_{001}-O_{031}+O_{301}-O_{331})^2\notag\\
&+(O_{002}-O_{032}+O_{302}-O_{332})^2].
\end{align}
This allows us to obtain the MF decoupled $\int d^2r\;U\hat{n}_{p_x,\uparrow}({\bf r})\hat{n}_{p_x,\downarrow}({\bf r})$ as
\begin{align}
\int d^2r\bigg\{\frac{U}{32}&[(O_{000}-O_{030}+O_{300}-O_{330})\notag\\
&\times\hat{\Psi}^{\dag}(\Sigma_{000}-\Sigma_{030}+\Sigma_{300}-\Sigma_{330})\hat{\Psi}\notag\\
&-(O_{003}-O_{033}+O_{303}-O_{333})\notag\\
&\times\hat{\Psi}^{\dag}(\Sigma_{003}-\Sigma_{033}+\Sigma_{303}-\Sigma_{333})\hat{\Psi}]\notag\\
-\frac{U}{32}&[(O_{001}-O_{031}+O_{301}-O_{331})\notag\\
&\times\hat{\Psi}^{\dag}(\Sigma_{001}-\Sigma_{031}+\Sigma_{301}-\Sigma_{331})\hat{\Psi}\notag\\
&+(O_{002}-O_{032}+O_{302}-O_{332})\notag\\
&\times\hat{\Psi}^{\dag}(\Sigma_{002}-\Sigma_{032}+\Sigma_{302}-\Sigma_{332})\hat{\Psi}]\bigg\}\notag\\
-S\frac{U}{64}&[(O_{000}-O_{030}+O_{300}-O_{330})^2\notag\\
&-(O_{003}-O_{033}+O_{303}-O_{333})^2]\notag\\
+S\frac{U}{64}&[(O_{001}-O_{031}+O_{301}-O_{331})^2\notag\\
&+(O_{002}-O_{032}+O_{302}-O_{332})^2].
\end{align}
By applying the same trick for the other interaction terms in Eq.~(\ref{Hint}), we finally find 64 MF undetermined parameters (summarized in Table.~\ref{T2}), and obtain the MF free energy per unit area at zero temperature as:
\be
F=\frac{1}{S}\int d^2r\;\hat\Psi^{\dag}(H_{(001)}+H_{int}^{MF})\hat\Psi-n\mu+E_{con},
\ee
where $H_{int}^{MF}$ is the MF decoupled interaction Hamiltonian for $\hat H_{int}$ and $E_{con}$ is the MF condensation energy. The ground state can then be determined by solving the set of coupled mean-field equations resulting from minimizing $F$ with respect to all possible order parameters.

\begin{table}[h]
\caption{\label{T2}Summary of MF parameters}
\begin{ruledtabular}
\begin{tabular}{m{2.5cm}|m{5.5cm}}
Type & MF parameters \\
\hline
Renormalized band parameters&$O_{030}$, 
$O_{022}$,\footnotemark[1] $O_{322}$,\footnote{In the absence of strain, if $O_{022}\neq O_{322}$, this is an order parameter for spontaneous mirror symmetry breaking.}
$O_{021}$,\footnotemark[2] $O_{321}$,\footnote{In the absence of strain, if $O_{021}\neq-O_{321}$, this is an order parameter for spontaneous mirror symmetry breaking.}
$O_{010}$,\footnotemark[3] $O_{310}$,\footnotemark[3] $O_{023}$,\footnotemark[3] $O_{323}$,\footnote{In the absence of strain, if one of $O_{010}$, $O_{310}$, $O_{023}$ and $O_{323}$ is nonzero, this would be an order parameter for spontaneous mirror symmetry breaking.}
$O_{300}$,\footnotemark[4] $O_{330}$,\footnote{In the absence of strain, if either $O_{300}$ or $O_{330}$ is nonzero, this would be an order parameter for spontaneous $C_4$ symmetry breaking.}
$O_{000}$ \\\hline
FM & $O_{002}$, $O_{032}$, $O_{001}$, $O_{031}$, $O_{003}$, $O_{033}$, $O_{013}$, $O_{313}$, $O_{020}$, $O_{320}$, $O_{011}$, $O_{311}$, $O_{012}$, $O_{312}$, $O_{303}$, $O_{333}$, $O_{301}$, $O_{331}$, $O_{302}$, $O_{332}$\\\hline
CDW & $O_{100}$, $O_{130}$, $O_{110}$, $O_{123}$, $O_{122}$, $O_{121}$, $O_{220}$, $O_{213}$, $O_{212}$, $O_{211}$, $O_{203}$, $O_{233}$, $O_{202}$, $O_{232}$, $O_{201}$, $O_{231}$\\\hline
SDW & $O_{103}$, $O_{133}$, $O_{102}$, $O_{132}$, $O_{101}$, $O_{131}$, $O_{113}$,  $O_{120}$, $O_{111}$, $O_{112}$, $O_{223}$, $O_{210}$, $O_{221}$, $O_{222}$, $O_{200}$, $O_{230}$
\end{tabular}
\end{ruledtabular}
\end{table}


\begin{thebibliography}{99}
\bibitem{Coulomb}
Importance of the Coulomb interaction for relativistic quasiparticles may be assessed approximately by the dimensionless interaction strength, $\alpha=\frac{e^2}{\epsilon v_F \hbar}$, where $\epsilon$ denotes the dielectric constant, and $v_F$ is the Fermi velocity. For graphene, $\epsilon\alpha\approx 2.2$, while for SnTe $\epsilon\alpha$ is even higher due to a smaller value of $v_F$.
\bibitem{long-range}
Effects of long-range Coulomb interaction should be examined further.
\end{thebibliography}
\end{document}